# Relationships between ligand binding sites, protein architecture and correlated paths of energy and conformational fluctuations

Burak Erman

Chemical and Biological Eng Dept

Koc University Istanbul Turkey

email: **berman@ku.edu.tr**

Short title: Energy and conformational fluctuations in proteins

# Abstract

Statistical thermodynamics basis of energy and residue position fluctuations is explained for native proteins. The protein and its surroundings are treated as a canonical system with emphasis on the effects of energy exchange between the two. Fluctuations of the energy are related to fluctuations of residue positions, which in turn are related to the connectivity matrix of the protein, thus establishing a connection between energy fluctuation pathways and protein architecture. The model gives the locations of hotspots for ligand binding, and identifies the pathways of energy conduction within the protein. Results are discussed in terms of two sets of models, the BPTI and twelve proteins that contain the PDZ domain. Possible use of the model for determining functionally similar domains in a diverse set of proteins is pointed out.

# 1. Introduction

The native protein in the cell is not an isolated system, but undergoes continual exchange of energy with its surroundings as a result of which the protein performs its function. The fluctuations of energy act as the driving potential behind several phenomena affecting the function of the protein, notably the spatial fluctuations of the residue positions. The latter are significant in native proteins at physiological temperatures, as evidenced by experimental B-factors. The amplitudes of fluctuations are spatially inhomogeneous, being different for different residues and exhibiting an inverse dependence on the number of spatial neighbors of each residue, thus establishing an important connection to the topology of the native state, or the contact map of the protein. The present study aims to explain the statistical thermodynamics basis of the relationship between energy and conformational fluctuations of the protein at the residue level. A deeper understanding of the thermostatistics of the protein allows us to answer, or at most consider, several questions that have recently been attracting the attention of investigators on the structural basis of energetic interactions upon ligand binding (Lockless and Ranganathan, 1999; Freire, 1999; Suel *et al.*, 2003). Specifically, we address the question of the partitioning of the instantaneous increment of energy from the surroundings among the various residues, and the relation of this partitioning to the topology of the three dimensional structure. The analysis shows, as will be discussed in detail in the results section, that the exchange of energy between the protein and the surroundings is not spatially isotropic, nor random. The surroundings may be water molecules as well as ligands, cell wall, DNA, proteins, etc. Certain residues, which we recently called the 'energy gates'(Tuzmen and Erman, 2011), are the hotspots that play major role in energy exchange with the surroundings. These are residues that can respond to the incoming energy and can share it with others in the protein, compared to several other residues at the surface that show negligible response to energy perturbations. This observation has immediate and important consequences regarding the ligand binding problem. The analysis also shows that the energy taken up from the surroundings does not diffuse randomly within the three dimensional structure, but rather follows specific paths which we recently called 'interaction pathways' (Haliloglu *et al.*, 2010; Tuzmen and Erman, 2011). The relation of this observation to allosteric interactions of proteins, on which a large number of studies now converge, is obvious (Monod *et al.*, 1965).

The interest in energy fluctuations in proteins is not new, of course, since it relates directly to the heat capacity, $C_P$, by the relation $C_P = \langle \Delta U^2 \rangle / kT^2$, which has been given most transparently in the work of Prabhu and Sharp (Prabhu and Sharp, 2005) on which we elaborate further in the models section below. Interest on energy fluctuations has been mostly limited to the study of changes in the heat capacity relating to events during folding or unfolding. Following the pioneering works of Cooper (Cooper, 1976) and Sturtevant (Sturtevant, 1977), several papers have been written on this subject and reviewed extensively in (Prabhu and Sharp, 2005). The sources of heat capacity changes with increasing temperature such as formation of cages of structured water around nonpolar groups, the breaking of hydrogen bonds and the increase of internal vibrational degrees of freedom have been studied in detail both theoretically and experimentally (Prabhu and Sharp, 2005; Sturtevant, 1977). In the present paper, we do not address the issue of heat capacity changes but instead focus on the partitioning of the fluctuations at the residue level, i.e., the relationship between energy fluctuations and the vibrational degrees of freedom of individual residues.

Several papers form the background and the underlying material for the present study which are briefly reviewed here. The terminology and the formulation of fluctuations in Chapter 19 of the classical book on thermostatistics by Callen (Callen, 1985) are adopted. This terminology was used for proteins previously (Oylumluoglu *et al.*, 2006, 2007) with emphasis on the electric field as the thermodynamic extensive variable and the protein dipole moment as the thermodynamic force. Electric field fluctuations are not considered here but can however be handled with the general expression of the probability function given in Appendix A. The present work is deeply motivated by the work of Piazza, de Los Rios and Sanejouand (Piazza *et al.*, 2005) who pointed out that the energy of a protein is dissipated to the environment only by surface atoms, and bulk atoms exchange energy with each other only. A simplified solution of the Fokker-Planck equation based on harmonic potentials showed that the spectrum of energy relaxation times originated from the coupling of the surface atoms to the solvent. The concept of spatial inhomogeneity that is needed to understand protein behavior originated in (Piazza *et al.*, 2005). In the present paper, residue positions are introduced as labels that carry information on the graph structure of the native protein, hence the structural inhomogeneity. The structural distribution of cooperative transitions and the identification of pathways of communication and their relation to energy fluctuations (Freire, 1999; Hilser *et al.*, 1998; Lockless and Ranganathan, 1999) can now be visualized within a thermostatistics framework. The statistical energy has already been recognized as a good indicator of coupling that determines the pathways of energy conduction within proteins (Lockless and Ranganathan, 1999; Suel *et al.*, 2003). The role of energy fluctuations on function has been studied from different points of view: The fact that interactions of proteins with their surroundings lead to diverse energy relaxation behavior of significant nature have been shown experimentally on myoglobin (Fenimore *et al.*, 2002; Frauenfelder *et al.*, 2002). Energy storage and transfer in the anisotropic three dimensional structure of a native protein are studied by cooling a surface residue and following the distribution of energy throughout the bulk (Piazza and Sanejouand, 2008, 2009). The flow and repartitioning of energy in proteins have been reviewed, thoroughly, by Leitner (Leitner, 2008) from the point of view of both (i) energy flow through channels containing the crucial residues, and (ii) collective normal modes of oscillations. The reader is referred to Leitner's review for the experimental and computational techniques that may be used to study energy fluctuations. On the computational side, the work of Fermi, Pasta, Ulam (Fermi *et al.*, 1955 ) have been the reference that influenced work on proteins: In an isolated system subject to harmonic potentials, energy given to a mode of the system remains there indefinitely, and it is only the anharmonicities in the potential that allows the energy to flow to different modes. It was later shown by the pioneering molecular dynamics simulations of Moritsugu, Miyashita and Kidera (Moritsugu *et al.*, 2000) that through a third order nonlinearity, the vibrational energy was transferred from a normal mode to a very few number of normal modes in a protein. When regarded as a canonical system, the modes of a protein, even in the absence of anharmonicities, may be excited by the surrounding liquid and a wide spectrum of energy relaxation may occur as pointed out before (Piazza *et al.*, 2005). In this respect, although anharmonicities are important, such as for forming discrete breathers (Juanico *et al.*, 2007; Piazza and Sanejouand, 2008) for example, they are not essential for the discussion of energy fluctuations in proteins. Finally, the large repertoire of papers on the relation of energy fluctuations on allosteric reactions of proteins, reviewed by Swain and Gierasch (Swain and Gierasch, 2006) is the relevant background to this work. Notably, the work of Ranganathan and collaborators (Lockless and Ranganathan, 1999; Suel *et al.*, 2003) has broadened our perspective on allosteric communication and protein architecture. The results and predictions of the fluctuation model of proteins presented here have close bearing to their work. Along similar lines, the model developed by Freire and collaborators (Freire, 1999) that emphasizes the propagation of binding interactions to remote sites in proteins depend heavily on the concept of energy fluctuations and their relation to protein structure.

A second aim of this paper is to recapitulate the statistical thermodynamics basis of the Gaussian Network Model (GNM) (Bahar *et al.*, 1997). This model has been derived with reference to the statistical mechanics of random Gaussian networks (Kloczkowski *et al.*, 1989), and has a partition function equivalent to that of a canonical system, but the significance of energy exchanges of the protein with its surroundings has not been visualized then. The latter is emphasized in the present study. In doing this, some of the familiar equations of the GNM are repeated, for the sole interest of explaining their connection to the canonical statistical mechanical system. More recent work on the thermostatistics of native proteins from our group elaborated on different aspects of the problem such as anharmonic probability distribution of residue fluctuations (Kabakcioglu *et al.*, 2010; Yogurtcu *et al.*, 2009), quasi-harmonic mode coupling (Gur and Erman, 2010), binding interactions (Haliloglu and Erman, 2009; Haliloglu *et al.*, 2008), and predicting interaction pathways (Haliloglu *et al.*, 2010; Tuzmen and Erman, 2011). These topics are unified under a thermostatistics formalism in the present paper, and it is hoped that this theoretical framework will allow for future improvements of the model.

The paper is organized as follows: In the section below, the thermodynamic variables are defined in the entropy representation for the model. The protein plus its surroundings form an isolated system, i.e., a canonical system. Then, the harmonic approximation is adopted and the relationship between the energy and residue fluctuations is given. The main finding of the paper, correlations of energy fluctuations is then presented in terms of two related matrices, the connectivity matrix of the protein as a graph and the gamma matrix of the GNM. The expression derived for energy fluctuation correlations establishes the relationship between energy fluctuations and protein architecture. Finally, the 'heat capacity of the distance between two residues' is introduced as a new concept which is a measure of the response of a pair of residues to energy fluctuations. All relevant thermostatistics information is summarized in six appendices. In the interest of giving a broad perspective to the thermostatistics of native proteins, the most general form of the probability distribution function for fluctuations is given in Appendix A, which is then simplified for use in this study. Some of the appendices contain information which is already well know but introduced here to eliminate cross-referencing, and some of the appendices contain details of mathematical derivations of the equations given in the text. The predictions of the model are compared, for proof of concept purposes, for two widely studied systems, (i) Bovine Pancreatic Trypsin Inhibitor, BPTI, and (ii) twelve proteins of the PDZ domain.

## 2. Material and Methods

***The model***: The model consists of a protein of $N$ residues embedded in surroundings of $N_S$ molecules. The protein and the surroundings form an isolated system. The pressure, temperature and total number of molecules are fixed. We assume that the protein is in a state of local energy minimum where small fluctuations away from mean positions are always restored back to the mean positions. The protein exchanges energy with the surroundings, resulting in fluctuations of the energies of the individual residues and of residue positions. In a general model, the volume of the protein also fluctuates, but we assume that changes in conformation leading to anisotropic fluctuations of shape are much larger than the changes in volume. We therefore assume the protein to be incompressible. We adopt the residue based coarse graining approximation, where atoms of each residue are centered at the corresponding alpha carbon, $C_i^\alpha$. As was pointed our recently, the residue level approximation yields the minimal set of thermodynamic variables needed to explain energy fluctuations in native proteins (Rader, 2010).

***Thermodynamic variables***: Each residue is identified by its position vector, $\boldsymbol{R}_i$, where the subscript i identifies the residue. Without the subscript, $\boldsymbol{R}$ represents the set of all position vectors of the $N$ residues. The thermodynamic extensive variables, the energy, $U(\boldsymbol{R})$, and entropy, $S(\boldsymbol{R})$, are functions of residue positions, with instantaneous values $\hat{U}(\hat{\boldsymbol{R}})$ and $\hat{S}(\hat{\boldsymbol{R}})$. We assign an entropy $S_i(\boldsymbol{R})$ and an energy $U_i(\boldsymbol{R})$ to each residue. The extensive nature of the entropy and energy requires that

$$U = \sum_{i=1}^{N} U_i \qquad S = \sum_{i=1}^{N} S_i \tag{1}$$

***The entropy representation of the protein***: The entropy of the ith residue may be written as a function of the energies of the constituent subsystems as

$$S_i = S_i(U_j(\boldsymbol{R})) \qquad j = 1, ...N \tag{2}$$

Equation (2) may be inverted to yield the energy of each residue

$$U_i = U_i(S_j(\boldsymbol{R})) \qquad j = 1, ...N \tag{3}$$

Using the first of equation (1), the total energy of the protein is written in the entropy representation as

$$U(S) = \sum_{i=1}^{N} U_i(S_j(\boldsymbol{R})) \qquad j = 1, ...N \tag{4}$$

***The system as a canonical ensemble***: The protein exchanges energy with its surroundings and the system constitutes a canonical ensemble. The probability $f(\hat{U}(\hat{\boldsymbol{R}}))$ that the protein has the instantaneous energy $\hat{U}(\hat{\boldsymbol{R}})$ follows from the general expression equation (A-1) as

$$f(\hat{U}(\hat{\boldsymbol{R}})) = \exp\left\{\frac{1}{kT}(A - \hat{U})\right\} \tag{5}$$

where, $A = U - TS$ is the Helmholtz free energy. The probability of a fluctuation $\Delta\hat{U} = \hat{U} - U$ of the energy is obtained from equation (5) as

$$f(\Delta\hat{U}) = \exp\left(-\frac{S}{k}\right)\exp\left(-\frac{1}{kT}\Delta\hat{U}\right) \tag{6}$$

***Relationship between energy and residue position fluctuations and the force-fluctuation relation***: The contributions to the fluctuating energy of a residue coming from the surroundings of that residue may be due to hydrogen bonding, Lenard-Jones type forces, dipolar coupling, electrostatic coupling, covalent bonding, etc. Whatever the source of the energy is, its fluctuations are coupled with the spatial fluctuations $\Delta\hat{\boldsymbol{R}}$ of the residues. In the simplest approximation, we assume that a residue i and j that are within the interaction distance of each other interact with a harmonic potential. The fluctuation of the energy $\Delta U_{ij}$ in this case is related to the residue position fluctuations $\Delta \hat{R}_i$ and $\Delta \hat{R}_j$ by (Hinsen, 1998)

$$\Delta U_{ij} = \left[ k_{ij} \cos^2 \left( \alpha_{ij} \right) \right] \left( \Delta \hat{R}_j - \Delta \hat{R}_i \right)^2 \tag{7}$$

where, $k_{ij}$ is the spring constant between residues i and j, $\alpha_{ij}$ is the angle between the vector $R_j - R_i$ and the vector $\Delta \hat{R}_j - \Delta \hat{R}_i$. The parameter $\cos^2 \left( \alpha_{ij} \right)$ is a function of instantaneous conformation, which we approximate by its average value $\left\langle \cos^2 \left( \alpha_{ij} \right) \right\rangle$ and lump into the spring constant term. Also, for the interest of simplicity, we assume that spring constants for all interacting pairs are equivalent. Thus, $\Delta U_{ij} = \gamma \left( \Delta \hat{R}_j - \Delta \hat{R}_i \right)^2$ where $\gamma$ is the equivalent spring constant for the system. We let this energy to partition equally between residues i and j. Summing up first over all the neighbors of residue i and then over all residues of the protein gives the total energy fluctuation

$$\Delta \hat{U} = \sum_i \Delta \hat{U}_i = \frac{1}{2} \gamma \sum_i \sum_j C_{ij} \left( \Delta \hat{R}_j - \Delta \hat{R}_i \right)^2 \tag{8}$$

The force $F_i$ on residue i when the residue is displaced from its mean position by $\Delta R_i$ is obtained from the energy as $F_i = \left( \dfrac{\partial \Delta U}{\partial \Delta R_i} \right)_{\Delta \hat{R}_j} = -\gamma \sum_j C_{ij} \left( \Delta \hat{R}_j - \Delta \hat{R}_i \right)$ which rearranges into

$$F_i = \Gamma_{ij} \Delta \hat{R}_j \tag{9}$$

where the matrix $\Gamma$ is the matrix of the GNM defined as

$$\Gamma_{ij} = \begin{cases} -\gamma C_{ij} & \text{if } i \neq j \\ \gamma \sum_{k \neq i} C_{ik} & \text{if } i = j \end{cases} \tag{10}$$

*The heat capacity and fluctuations of energy and entropy*: The heat capacity at constant pressure is $C_P = \frac{\partial H}{\partial T} = \frac{\partial U}{\partial T}$, where the second equality follows from the incompressibility assumption of the model. For the canonical ensemble, equation (B-1) gives

$$C_P = \frac{\partial U}{\partial T} = \frac{1}{kT^2}\left\langle \left(\Delta \hat{U}\right)^2 \right\rangle = \frac{1}{kT^2}\left\langle \sum_i \sum_j \Delta \hat{U}_i \Delta \hat{U}_j \right\rangle \tag{11}$$

Substitution from equation (8) leads to

$$C_P = \frac{1}{4}\frac{\gamma^2}{kT^2}\sum_i \sum_k \sum_j \sum_l C_{ij}C_{kl}\left\langle \left(\Delta\hat{\boldsymbol{R}}_i - \Delta\hat{\boldsymbol{R}}_j\right)^2 \left(\Delta\hat{\boldsymbol{R}}_k - \Delta\hat{\boldsymbol{R}}_l\right)^2 \right\rangle \tag{12}$$

An equivalent definition of the heat capacity is $C_P = T\frac{\partial S}{\partial T}$. Applying equation (B-1) to the right hand side of this expression, we obtain

$$C_P = T\frac{\partial S}{\partial T} = \frac{1}{kT}\left\langle \Delta\hat{S}\Delta\hat{U} \right\rangle = \frac{1}{kT}\left\langle \sum_i \sum_j \Delta\hat{S}_i \Delta\hat{U}_j \right\rangle \tag{13}$$

For the canonical ensemble, unlike the energy and the entropy, the free energy of the system does not fluctuate, as shown in Appendix C. Thus, $\left\langle \left(\Delta A\right)^2 \right\rangle = 0$ and we obtain

$\left\langle \left(\Delta\hat{A}\right)^2 \right\rangle = \left\langle \left(\Delta\hat{U} - T\Delta\hat{S}\right)^2 \right\rangle = 0$. Expanding the right hand side, and substituting from equations (11) and (13), we obtain the relationship between entropy and energy fluctuations.

$$\left\langle \left(\Delta\hat{S}\right)^2 \right\rangle = T^{-2}\left\langle \left(\Delta\hat{U}\right)^2 \right\rangle \tag{14}$$

Thus, entropy fluctuations are not independent of energy fluctuations.

*Correlations of energy fluctuations*: We now examine in more detail the contributions to the heat capacity from interactions of residue pairs

$$\left\langle \Delta\hat{U}_i \Delta\hat{U}_k \right\rangle = \frac{1}{4}\gamma^2 \sum_j \sum_l C_{ij}C_{kl}\left\langle \left(\Delta\hat{\boldsymbol{R}}_i - \Delta\hat{\boldsymbol{R}}_j\right)^2 \left(\Delta\hat{\boldsymbol{R}}_k - \Delta\hat{\boldsymbol{R}}_l\right)^2 \right\rangle \tag{15}$$

Expanding the terms in the brackets, and performing the indicated averages as outlined in Appendix E, leads to the final expression for correlations of the energy fluctuations

$$\langle \Delta \hat{U}_i \Delta \hat{U}_k \rangle = \frac{1}{4}(kT)^2 \sum_j \sum_l C_{ij} C_{kl} \Big[ 2\big((\Gamma_{ik}^{-1})^2 + (\Gamma_{il}^{-1})^2 + (\Gamma_{jk}^{-1})^2 + (\Gamma_{jl}^{-1})^2\big)$$

$$+ \Gamma_{ii}^{-1}\Gamma_{kk}^{-1} + \Gamma_{ii}^{-1}\Gamma_{ll}^{-1} + \Gamma_{jj}^{-1}\Gamma_{kk}^{-1} + \Gamma_{jj}^{-1}\Gamma_{ll}^{-1}$$

$$-4\big(\Gamma_{il}^{-1}\Gamma_{ik}^{-1} + \Gamma_{jl}^{-1}\Gamma_{jk}^{-1} + \Gamma_{ik}^{-1}\Gamma_{jk}^{-1} + 2\Gamma_{il}^{-1}\Gamma_{jl}^{-1}\big) \quad (16)$$

$$-2\big(\Gamma_{ii}^{-1}\Gamma_{kl}^{-1} + \Gamma_{jj}^{-1}\Gamma_{kl}^{-1} + \Gamma_{kk}^{-1}\Gamma_{ij}^{-1} + \Gamma_{ll}^{-1}\Gamma_{ij}^{-1}\big)$$

$$+4\big(\Gamma_{ij}^{-1}\Gamma_{kl}^{-1} + \Gamma_{ik}^{-1}\Gamma_{jl}^{-1} + \Gamma_{il}^{-1}\Gamma_{kj}^{-1}\big) \Big]$$

This is the major result of the present paper.

At a fixed temperature, the correlation given by equation (16) is proportional to the contribution of the interaction of residues i and j to the heat capacity. Summing equation (16) over the index k leads to the energetic interaction, $\Delta U_i$ of residue i with the remaining residues of the protein:

$$\Delta U_i = \sum_k \langle \Delta \hat{U}_i \Delta \hat{U}_k \rangle \quad (17)$$

***The temperature coefficient of the mean squared distance between two residues***: We now investigate the temperature coefficient $\dfrac{\partial \langle \Delta R_{ij}^2 \rangle}{\partial T}$ of the mean squared distance between residues i and j. This quantity shows the relative amount of the total energy observed by the specific inter-residue interaction. The relationship of the derivative to energy fluctuations follows from equation (B-1) as $\dfrac{\partial \langle \Delta R_{ij}^2 \rangle}{\partial T} = \dfrac{1}{kT^2} \langle \Delta \hat{R}_{ij}^2 \Delta \hat{U} \rangle$. In order to further characterize this relation for harmonic interactions, we evaluate the more general term $\langle \Delta \hat{R}_i^2 \Delta \hat{R}_j^2 \Delta \hat{U} \rangle$ which previously was derived in Reference (Haliloglu *et al.*, 2010) and is briefly summarized in Appendix F. The resulting expression is

$$\frac{\partial \langle \Delta R_{ij}^2 \rangle}{\partial T} = \frac{1}{kT^2} \langle \Delta \hat{R}_{ij}^2 \Delta \hat{U} \rangle = k \left( \Gamma_{ii}^{-1} - 2\Gamma_{ij}^{-1} + \Gamma_{jj}^{-1} \right) \tag{18}$$

Summing over the index j in equation (18) leads to the total correlation, $C_{T,i}$, of residue i with all other residues of the protein:

$$C_{T,i} = k \sum_j \left( \Gamma_{ii}^{-1} - 2\Gamma_{ij}^{-1} + \Gamma_{jj}^{-1} \right) \tag{19}$$

## 3. Results and Discussion

We now apply the predictions of equations (16) and (17) to the analysis of energy interactions of two widely studied systems, (i) Bovine Pancreatic Trypsin Inhibitor, BPTI, and (ii) the PDZ domain. In all calculations, the cutoff distance is taken as 7.0 Å. Our interest is in localized motions that identify specific residues. For this reason, we concentrate on the large eigenvalue end of the spectrum of the gamma matrix. We retain the largest five eigenvalues of the gamma matrix in calculating $\Gamma^{-1}$. Contributions from eigenvalues beyond the fifth do not contribute much to local events as our calculations show. In previous studies, we concentrated mostly on the largest eigenvalue (Haliloglu and Erman, 2009; Haliloglu *et al.*, 2010; Haliloglu *et al.*, 2008; Tuzmen and Erman, 2011) which accounted for the majority of events associated with the correlations that we studied. Using five largest eigenvalues now gives a consistent picture of the fine details of the energy-structure relations as we discuss below.

(i) BPTI. Calculations are performed using two pdb structures, 1BPI.pdb and 4PTI.pdb that are obtained in the presence and absence of the ligand, respectively. The results of equation (16) are presented in a contour diagram in figure 1 for 1BPI. The dark regions show the large values of the correlation $\langle \Delta \hat{U}_i \Delta \hat{U}_j \rangle$ of energy fluctuations between residues i and j. According to the results, ARG20 and TYR21 are correlated with TYR35 and CYS51. TYR35 is correlated with CYS51.

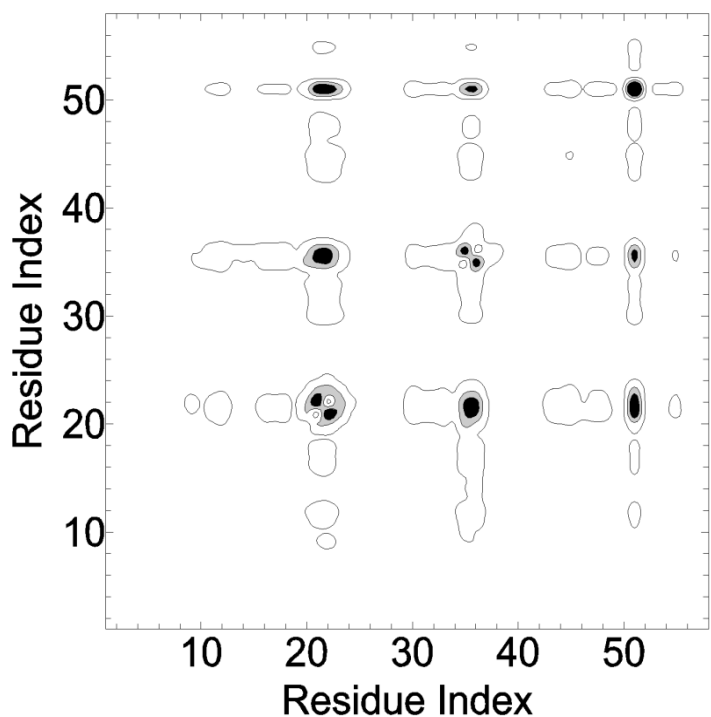

Figure 1. Contour diagram for $\left\langle \Delta\hat{U}_i \Delta\hat{U}_j \right\rangle$ showing the important correlations between residues ARG20, TYR21, TYR35 and CYS51.

The results obtained from equation (17) are shown in figure 2. The heavy line is obtained by using the PDB file 1BPI whose crystal structure is obtained in the presence of the phosphate group that binds to ARG20 and TYR35. The two peaks identify the binding site residues for the ligand. The light curve is obtained by using the PDB file 4PTI which is crystallized in the absence of the ligand. The curve obtained is essentially the same as that for the liganded protein. This shows that the information for binding of the ligand at the specified position is already present in the apo form of the protein. CYS30 and CYS51 make a disulfide bridge. Figure 1 shows that there is some but not a strong correlation between these two residues. The two other disulfide bridges, 14-38 and 5-55 do not appear to be interacting energetically according to the present model. Neither of these two is on the interaction pathway of this protein which we define below. PHE45 appears as a small peak in the figure. From figure 1, we see that PHE45 correlates with ARG20, TYR35 and CYS51. PHE45-ARG20 and PHE45-CYS51 are contact interactions whereas the correlations of PHE45-TYR35 are long distance correlations as seen from figure 3 below.

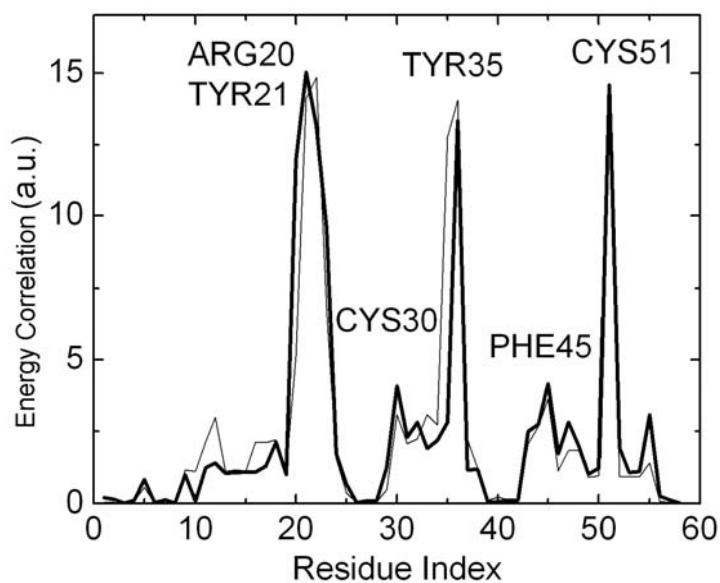

Figure 2. Results of calculations based on equation (17) showing the energetic correlation of a given residue. The correlations are given in arbitrary units. the light curve is obtained from the apo form of BPTI, the solid curve is with the ligand.

Figures 3 a and b show the interaction path that is obtained with the present study. The ligand PO4 is shown in red. It makes a contact with ARG20 ans TYR35 as shown in figure 3b. The path through which energy is transmitted in the protein is summarized by figure 3b. The path starts with ARG20 and TYR35, goes through TYR21 which makes a distance of 2.8 Å with PHE45, which in turn makes a distance of 3.96 Å with CYS51. Finally, CYS51 makes a sulfide bridge with CYS30. Thus, the calculations show that the interaction pathway in BPTI is through ARG20-TYR21-PGHE45-CYS51-CYS30. TYR35 AND ARG20 form the energy gate at one end of the pathway, CYS30 is at the other end.

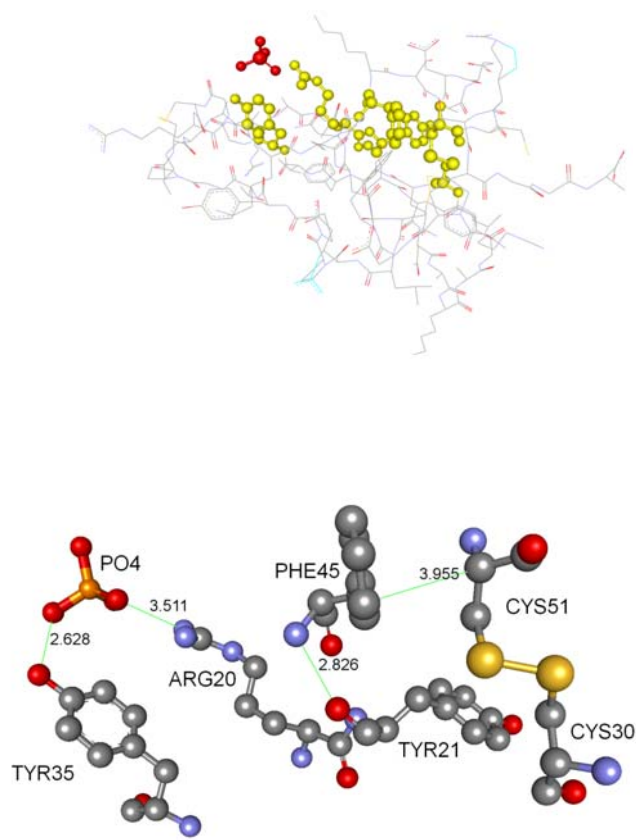

Figure 3a. The position of the interaction pathway in PBTI. 3b. enlarged view of the residues along the interaction pathway.

(ii) The PDZ system. The second application of the energy fluctuation model is on a set of twelve proteins with PDZ domain. Their Protein Data Bank identities are: 3I4W, 3NGH, 3JXT, 3QIK, 2KQF, 2W7R, 3PS4, 3QJM, 2KAW, 2KOJ, 3KHF, 2KG2. The correlations of energy fluctuations are calculated from equation (17) and the results are presented in the twelve panels of figure 6. All of the proteins in this group exhibit six characteristic peaks that are similarly located on the primary sequence of each protein. We consider 3I4W in detail here. In figure 4, results of calculations based on equation (17) are presented where energy correlations of residues identified by the residue index along the abscissa are presented in arbitrary units. For uniformity, residue indices are numbered from 1 to N and do not correspond to the actual residue numbers given in the data bank files. Six major peaks are observed, numbered from left to right in the figure. The residues corresponding to the six peaks are: ILE316 at peak 1, PHE325,ASN326 and ILE327 at peak 2, ILE336 at peak 3, PRO346 at peak 4, ASP357 at peak 5, GLN391 at peak 6. The corresponding structures are shown in figure 5 in 3-d.

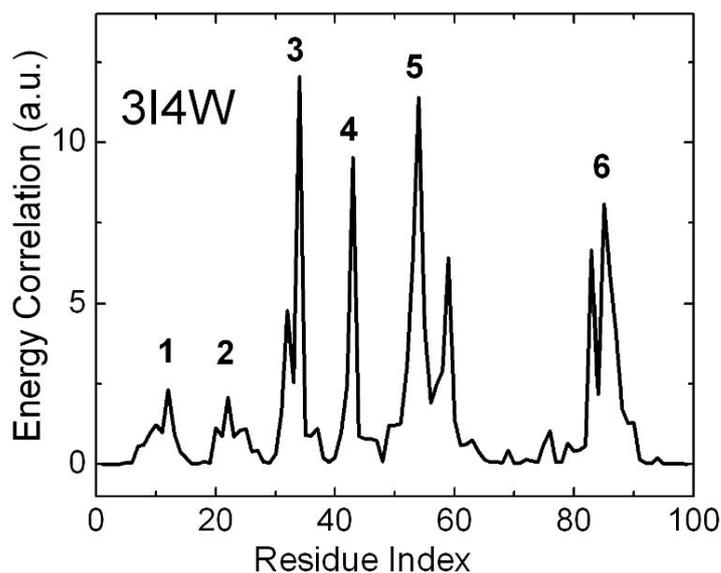

Figure 4. The six peaks of energy correlation determining the interaction path in the PDZ domain for the protein 3I4W.

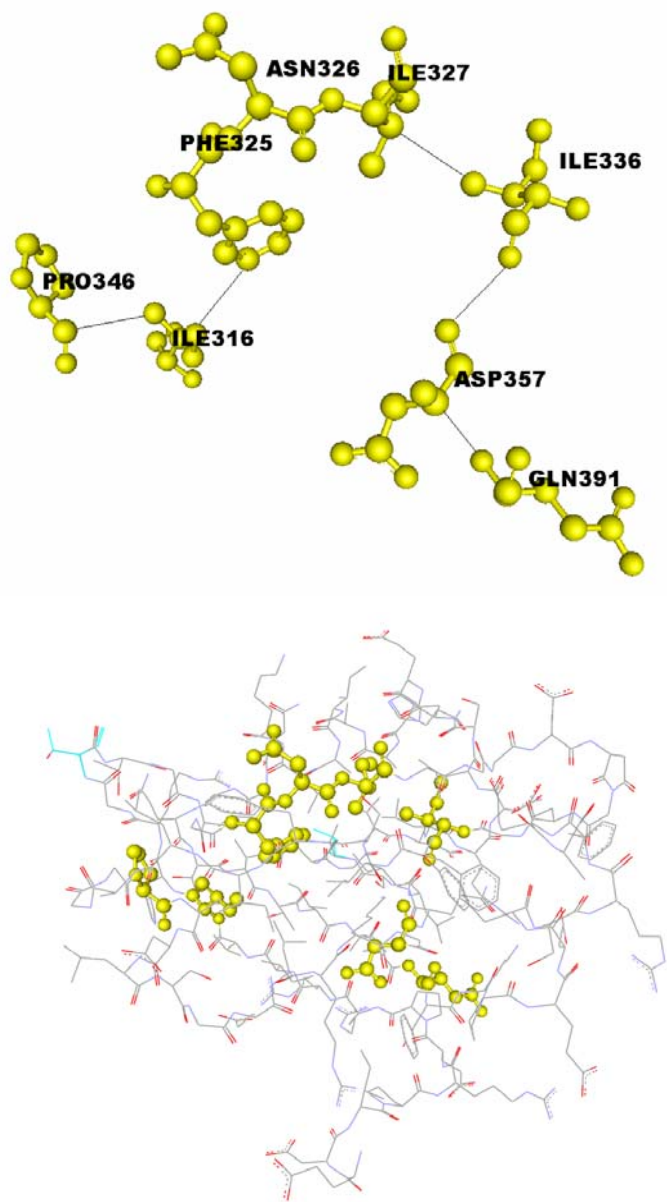

Figure 5a. The three dimensional structure of 3I4W. The residues along the interaction path are shown in yellow. 5b. The path members and the shortest distance between them. The shortest distances shown by the black line are all less than 4Å.

In figure 6, the energy correlation peaks are shown for all of the twelve proteins that we investigated. In all of the proteins, the characteristic six peaks are observable. In some cases, there are shifts in the locations of the peaks and in their amplitudes due to differences in the numbering of the residues and due to effects from the diverse neighborhoods of the proteins, but in all cases the characteristic peak structure is recognizable and are in general agreement with the patterns suggested for similar systems by Lockless and Ranganathan (Lockless and Ranganathan, 1999).

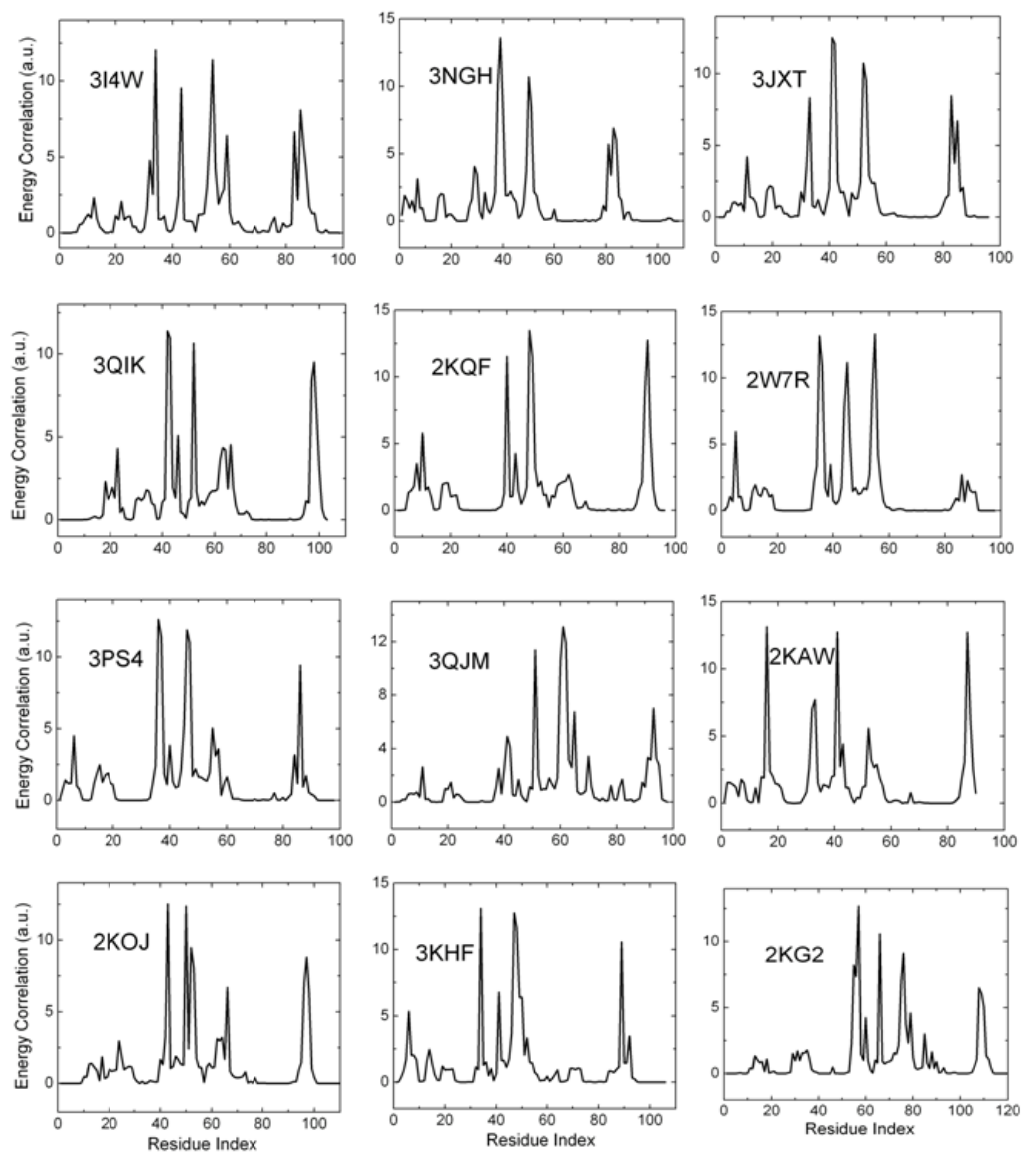

Figure 6. The energy correlation diagrams for the twelve proteins that contain the PDZ domain.

The primary aim of this paper was to establish a connection between a statistical mechanical description of proteins and energetic interaction that define their function. The structural basis of energetic perturbations and fluctuations entered into the model in a simple way through the connectivity matrix, thus establishing the relation of processes to protein architecture. In this sense, the present study outlines the conceptual framework for the energy processes in proteins. It can be improved in several directions. Although a harmonic potanetial is used for the examples, the general thermostatistics treatment is not confined to harmonic interactions. Anharmonicities may be introduced by the use of equation (D-1) by a suitable choice of force-fluctuation equations of state. The two examples presented above are intended for a proof of principle. More detailed analysis is needed to demonstrate the capabilities of the model. Our recent work (Tuzmen and Erman, 2011) on a related formulation, applied to 24 benchmark proteins are encouraging in this respect. The cutoff distance of 7.0 Å seems to be arbitrary, as has been the case in all previous studies of the GNM. A scaling study of the cutoff distance based on 4810 non-redundant structures obtained from the following web address (http://dunbrack.fccc.edu/Guoli/culledpdb) suggested that 7.0 Å is a reasonable value for the cutoff distance. Keeping only five largest eigenvalues in the formulation to represent localized events also seems arbitrary and further work on this is needed. For the present paper, five eigenvalues were necessary and sufficient to reflect the basic features of the PDZ domain, for example. Fewer eigenvalues reflected some but not all features, and a larger number of eigenvalues added to the redundancy in the results. The energy interaction pathways presented in figures 3 and 5 do not lie on a straight line but are rather of fractal nature. There should be a deep relation between the fractal nature of the pathways and protein function, which we cannot see now, but is definitely worthy of further examination. Finally, it seems possible to apply the model to the determination of unknown domains of interaction in a diverse set of proteins, simply by searching similar peaks, as it is the case with the PDZ domain.

**Acknowledgments:** Support by the Turkish Academy of Sciences is gratefully acknowledged.

# 4. Appendices

**Appendix A.** The probability distribution of fluctuations. (From Callen (Callen, 1985))

The general form of the probability function for the instantaneous values $\hat{X}$ of the thermodynamic variables is

$$f(\hat{X}) = \exp\left\{-\frac{1}{k}S[\mathcal{F}_0,...,\mathcal{F}_m] - \frac{1}{k}\sum \mathcal{F}_i \hat{X}_i\right\} \tag{A-1}$$

Here, $\hat{X}_i$ are the instantaneous values of the parameters of the model, which in general could be the energy and position of each residue, their volumes, electric field acting on each residue, etc. $\mathcal{F}_i$ are the corresponding entropic variables, such as $\frac{1}{T}$, $-\frac{F_i}{T}$, $\frac{P}{T}$, $-\frac{E_i}{T}$, etc. with showing the constant force acting on residue i, $P$ the pressure and $E_i$ the electric field on residue i. $S[\mathcal{F}_0,...,\mathcal{F}_m]$ is the general Massieu transform of the entropy with respect to its arguments. In the present paper we assume that the protein exchanges energy with its surroundings only, no constant forces and electric field acts on it, and therefore $\mathcal{F}_0 = \frac{1}{T}$, $\hat{X}_i = -\frac{\hat{U}_i}{T}$, and $S\left[\frac{1}{T}\right] = S - \frac{1}{T}U$. The resulting canonical form is equation (5) of the text.

**Appendix B**. Temperature derivative of a thermodynamic variable (From Prabhu and Sharp (Prabhu and Sharp, 2005))

For the canonical ensemble, using the probability function given by equation (7), the following relation is obtained

$$\frac{d\langle \hat{X}\rangle}{dT} = \frac{1}{kT^2}\langle \hat{X}\hat{U}\rangle - \langle \hat{X}\rangle\langle \hat{U}\rangle = \frac{1}{kT^2}\langle \Delta\hat{X}\Delta\hat{U}\rangle \tag{B-1}$$

**Appendix C**. The expression $\langle (\Delta A)^2 \rangle = 0$ (From Meirovitch (Meirovitch, 1999))

Writing the Helmholtz free energy, $A$ as $A = \sum_i f_i A_i = \sum_i f_i (U_i + kT \ln f_i)$ and substituting for the probability from equation (5), i.e., $f_i = \exp\left\{-\frac{1}{kT}\hat{U}\right\}/Z$ gives $A_i = -kT \ln Z$ which is independent of the instantaneous states of the protein, hence does not fluctuate.

**Appendix D.** Fluctuation recursion relation (From Callen (Callen, 1985))

For any general system, the correlation of the fluctuations of any two thermodynamic variables $\Delta \hat{X}_j$ and $\Delta \hat{X}_k$ is given by the derivative

$$\langle \Delta \hat{X}_j \Delta \hat{X}_k \rangle = -k \left(\frac{\partial X_j}{\partial \mathcal{F}_k}\right)_{\mathcal{F}_1 \ldots \mathcal{F}_{k-1}, \mathcal{F}_{k+1} \ldots} = kT \left(\frac{\partial X_j}{\partial F_k}\right)_{F_1 \ldots F_{k-1}, F_{k+1} \ldots} \quad \text{(D-1)}$$

Repeated use of this expression for any fluctuating function $\hat{\phi}$ gives

$$\langle \hat{\phi} \Delta \hat{X}_k \rangle = -k \left\langle \hat{\phi} \frac{\partial f}{\partial \mathcal{F}_k} \right\rangle = -k \frac{\partial \langle \hat{\phi} \rangle}{\partial \mathcal{F}_k} + k \left\langle \frac{\partial \hat{\phi}}{\partial \mathcal{F}_k} \right\rangle \quad \text{(D-2)}$$

The thermostatistical basis of the GNM equation

$$\langle \Delta \hat{R}_j \Delta R_k^T \rangle = kT \left(\Gamma^{-1}\right)_{jk} \quad \text{(D-3)}$$

is obtained from equation (D-1), by choosing $F_i = \Gamma_{ij} \Delta \hat{R}_j$

**Appendix E**: Fourth order correlations of residue fluctuations

Expanding the right hand side of equation (16) leads to

$$\langle \Delta \hat{U}_i \Delta \hat{U}_k \rangle = \frac{1}{4} \gamma^2 \sum_j \sum_l C_{ij} C_{kl} \left[ \langle \Delta \hat{R}_i^2 \Delta \hat{R}_k^2 \rangle + \langle \Delta \hat{R}_i^2 \Delta \hat{R}_l^2 \rangle + \langle \Delta \hat{R}_j^2 \Delta \hat{R}_k^2 \rangle + \langle \Delta \hat{R}_j^2 \Delta \hat{R}_l^2 \rangle \right.$$

$$-2 \left(\langle \Delta R_i^2 \Delta R_k \Delta R_l \rangle\right) + \left(\langle \Delta R_j^2 \Delta R_k \Delta R_l \rangle\right) + \left(\langle \Delta R_k^2 \Delta R_i \Delta R_j \rangle\right) + \left(\langle \Delta R_l^2 \Delta R_i \Delta R_j \rangle\right) \quad \text{(E-1)}$$

$$\left. +4 \langle \Delta R_i \Delta R_j \Delta R_k \Delta R_l \rangle \right]$$

Using the relation D-1 and D-2, and the relation $F_i = \Gamma_{ij}\Delta R_j$ the higher moments of fluctuations are replaced by the products of the matrix $\Gamma$ as follows:

$$\langle \Delta \hat{R}_i^2 \Delta \hat{R}_k^2 \rangle = \left[ 2\left(\Gamma_{ik}^{-1}\right)^2 + \Gamma_{ii}^{-1}\Gamma_{kk}^{-1} \right]\frac{(kT)^2}{\gamma^2}$$

$$\langle \Delta \hat{R}_i^2 \Delta \hat{R}_l^2 \rangle = \left[ 2\left(\Gamma_{il}^{-1}\right)^2 + \Gamma_{ii}^{-1}\Gamma_{ll}^{-1} \right]\frac{(kT)^2}{\gamma^2}$$

$$\langle \Delta \hat{R}_j^2 \Delta \hat{R}_k^2 \rangle = \left[ 2\left(\Gamma_{jk}^{-1}\right)^2 + \Gamma_{jj}^{-1}\Gamma_{kk}^{-1} \right]\frac{(kT)^2}{\gamma^2}$$

$$\langle \Delta \hat{R}_j^2 \Delta \hat{R}_l^2 \rangle = \left[ 2\left(\Gamma_{jl}^{-1}\right)^2 + \Gamma_{jj}^{-1}\Gamma_{ll}^{-1} \right]\frac{(kT)^2}{\gamma^2} \tag{E-2}$$

$$\langle \Delta R_i^2 \Delta R_k \Delta R_l \rangle = \left[ 2\Gamma_{il}^{-1}\Gamma_{ik}^{-1} + \Gamma_{ii}^{-1}\Gamma_{kl}^{-1} \right]\frac{(kT)^2}{\gamma^2}$$

$$\langle \Delta R_j^2 \Delta R_k \Delta R_l \rangle = \left[ 2\Gamma_{jl}^{-1}\Gamma_{jk}^{-1} + \Gamma_{jj}^{-1}\Gamma_{kl}^{-1} \right]\frac{(kT)^2}{\gamma^2}$$

$$\langle \Delta R_k^2 \Delta R_i \Delta R_j \rangle = \left[ 2\Gamma_{ik}^{-1}\Gamma_{jk}^{-1} + \Gamma_{kk}^{-1}\Gamma_{ij}^{-1} \right]\frac{(kT)^2}{\gamma^2}$$

$$\langle \Delta R_l^2 \Delta R_i \Delta R_j \rangle = \left[ 2\Gamma_{il}^{-1}\Gamma_{jl}^{-1} + \Gamma_{ll}^{-1}\Gamma_{ij}^{-1} \right]\frac{(kT)^2}{\gamma^2}$$

$$\langle \Delta R_i \Delta R_j \Delta R_k \Delta R_l \rangle = \left[ \Gamma_{ij}^{-1}\Gamma_{kl}^{-1} + \Gamma_{ik}^{-1}\Gamma_{jl}^{-1} + \Gamma_{il}^{-1}\Gamma_{kj}^{-1} \right]\frac{(kT)^2}{\gamma^2}$$

**Appendix F**: Temperature coefficient of $\langle \Delta R_{ij}^2 \rangle$

Starting with the definition of energy fluctuation (Haliloglu *et al.*, 2010; Tuzmen and Erman, 2011)

$$\Delta U = F_k \Delta R_k = F_k F_j \left(\Gamma^{-1}\right)_{jk} \tag{F-1}$$

and applying the expressions D-1 and D-2, we obtain

$$\left\langle \Delta U \Delta \boldsymbol{R}_i \Delta \boldsymbol{R}_j^T \right\rangle = kT \left\langle \Delta \boldsymbol{R}_i \Delta \boldsymbol{R}_j^T \right\rangle = (kT)^2 \left(\Gamma^{-1}\right)_{jk} \tag{F-2}$$

Using the relation $\left\langle \Delta \hat{R}_{ij}^2 \right\rangle = kT \left( \boldsymbol{\Gamma}_{ii}^{-1} - 2\boldsymbol{\Gamma}_{ij}^{-1} + \boldsymbol{\Gamma}_{jj}^{-1} \right)$ and F-2 leads to equation (18).